\documentclass[10pt]{llncs}

\usepackage[utf8]{inputenc}
\usepackage[T1]{fontenc}
 \usepackage{caption}
\usepackage[fleqn]{amsmath}
\usepackage{amssymb,amsfonts,comment}
\usepackage{multirow,booktabs,todonotes}
\usepackage{algorithm,algorithmicx} 
\usepackage[noend]{algpseudocode}
\usepackage{varwidth,enumitem,tikz,tikz-3dplot} 
\usepackage[british]{babel}
\usepackage[final,activate=true,kerning=true,protrusion=true]{microtype}

\usepackage{zi4,tikz}
\usetikzlibrary{arrows,automata,positioning, shapes.symbols,shapes.callouts,patterns}
\usetikzlibrary{calc,shapes.geometric} 
\usetikzlibrary{fit,arrows.meta}

\usepackage[pdfpagelabels=true]{hyperref}

\usepackage{cleveref,xcolor}
\hypersetup{
	linktoc = page,
	pdfpagemode = UseNone,
	colorlinks,
	linkcolor={red!50!black},
	citecolor={blue!50!black},
	urlcolor={blue!80!black}
}

\usepackage{pgfplots,adjustbox,csquotes}
\usepackage[tikz]{bclogo}
\usepackage{xurl}

\usepackage{array,booktabs}

\begin{document}

\newcommand*\circled[1]{\tikz[baseline=(char.base)]{
            \node[shape=circle,draw,inner sep=1pt] (char) {\scriptsize #1};}}

\title{Explaining the Entombed Algorithm}

%\iffalse 

\author{Leon Mächler and David Naccache} 

\institute{ENS, CNRS, PSL Research University\\
  Département d'informatique, \'Ecole normale supérieure, Paris, France\\
\url{leon-philipp.machler@ens.fr} \\
\url{david.naccache@ens.fr}
}

%\else
%\author{(Submission to wherever)}
%\institute{}
%\fi

\maketitle         
\selectlanguage{british}
\begin{abstract} In \cite{entombed}, John Aycock and Tara Copplestone pose an open question, namely the explanation of the mysterious lookup table used in the Entombed Game's Algorithm for two dimensional maze generation. The question attracted media attention (BBC etc.) and was  open until today. This paper answers this question, explains the algorithm and even extends it to three dimensions.

\end{abstract}

\section{Introduction}

When Aycock and Copplestone reverse engineered the code of the 1982 Atari Game ``Entombed'' they found an inexplicable algorithm that consistently creates novel playable two dimensional mazes \cite{entombed}. The algorithm is interesting due to the fact that it's decisions are only based on local information which makes it fast and efficient. The local information is the 5 bit context of blocks surrounding the current position. For for each of the 32 possible scenari, a decision is predefined in a lookup table. It was unclear to Aycock and Copplestone how this lookup table was created, which they posed as an open question. 

\begin{figure}
    \centering
    \includegraphics[width=0.6\textwidth]{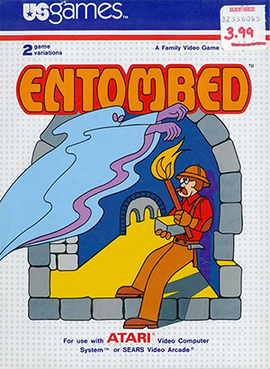}
    \caption{The game's coverart}
    \label{fig:pic1}
\end{figure}

\begin{figure}
    \centering
    \includegraphics[width=0.6\textwidth]{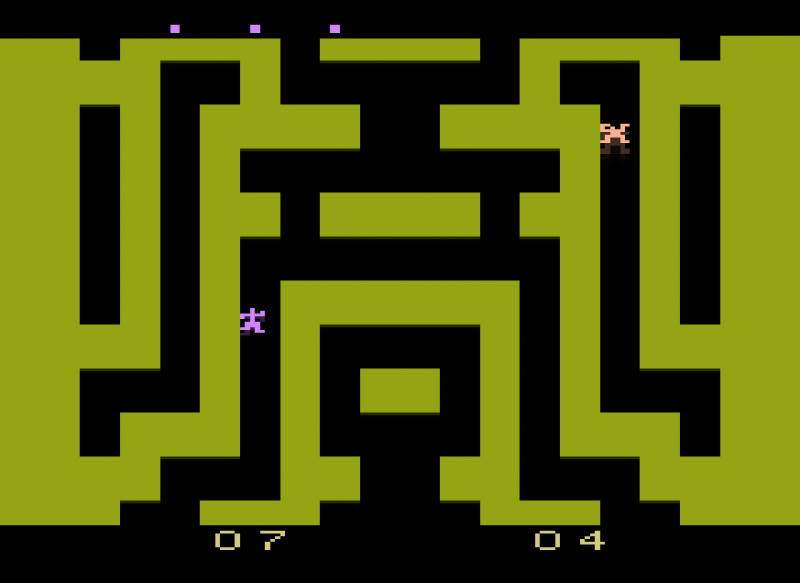}
    \caption{A typical maze generated by the algorithm}
    \label{fig:pic2}
\end{figure}

As of April 2021, the question was on Wikipedia's ``List of unsolved problems in computer science''\footnote{
\url{en.wikipedia.org/wiki/List_of_unsolved_problems_in_computer_science}}:

\begin{displayquote}
``\textsl{What is the algorithm for the lookup table that consistently generates playable mazes in the 1982 Atari 2600 game Entombed merely from the values of the five pixels adjacent to the next ones to be generated?}''
\end{displayquote}

Whereas the Entombed Wikipedia page\footnote{
\url{en.wikipedia.org/wiki/Entombed_(Atari_2600)}} states:

\begin{displayquote}
``\textsl{The mechanics of how Entombed generated its mazes have been the subject of academic research and some legend, as the maze data itself, if stored directly, was too large to fit within the hardware limitations of the console, even with the left/right symmetry of the mazes. Researchers evaluated the game's ROM and discovered that the mazes were generated on-the-fly by the game using the state of five adjacent squares of the maze (wall or open) already generated to determine the next part of the maze through a lookup table, including potentially a random state. Sometimes the table generates mazes that would be unsolvable without the "make-break" item. The researchers spoke to Sidley, who said the algorithm came from another unnamed programmer, but Sidley himself could not decipher why it worked. Sidley said to the researchers of this programmer, "He told me it came upon him when he was drunk and whacked out of his brain." Later research, however, suggests that although the original programmer and his mathematical collaborator did devise the algorithm at a bar, they may have suggested or encouraged the drunken blackout story merely to avoid having to explain or assign intellectual property rights for the algorithm.}''
\end{displayquote}

The question has attracted media attention\footnote{
\url{www.bbc.com/future/article/20190919-the-maze-puzzle-hidden-within-an-early-video-game}}\footnote{\url{www.wnycstudios.org/podcasts/tnyradiohour/segments/unearthing-entombed}}\footnote{\url{medium.com/codex/random-maze-from-entombed-8bb3b34e8f9b}}\footnote{\url{hackaday.com/2019/09/30/emtombed-secrets-partially-unearthed-as-researchers-dissect-clever-maze-generating-algorithm/}}\footnote{\url{www.techspot.com/news/85622-nobody-sure-what-makes-atari-2600-game-entombed.html}} and resulted in a dedicated Reddit thread\footnote{\url{www.reddit.com/r/math/comments/d8bgbu/a_mysterious_maze_algorithm/}}. This paper shows that all of the choices encoded in the lookup table are consistent with maintaining three simple invariants. When neither choice would violate an invariant a random choice is made. We then extend the method to three dimensions.

\section{The Algorithm}

The goal of the algorithm is to create random fixed-width (possibly infinite length) bi-dimensional mazes. This is achieved by successively creating line after line in the maze where each line is built block by block. A block is represented by one bit that can be interpreted as either $1$ (wall) or $0$ (path). The choice for each block is based solely on the immediately surrounding five blocks (two to the left, three above). For all 32 possible neighboring block combinations, the decision is predefined in a lookup table and can either be $1,0$ or $\mbox{\textsf{random}}$. When the neighboring blocks fall outside of the maze, either a random value or a predefined value is chosen instead. The decisions' hyper-locality allows for very cheap computational cost while still producing a vast plurality of possible output mazes. When this algorithm was discovered by Aycock and Copplestone in the code of an old Atari game called ``Entombed'', they could not explain it and left its clarification as an open question.

\begin{algorithm}[H]
\captionsetup{labelfont={sc,bf}, labelsep=newline}
  \caption{Entombed Maze Algorithm}
\begin{algorithmic} 
\Require The lookup table $L$, maze dimensions $X$ and $Y$
  \Ensure A binary maze $M$, where $M_{i,j} = 1$ iff a wall is at coordinate $i,j$. 

  \For{$i,j \leq X,Y$} 
  
  \State $a \gets M_{i,j-2}$
  \State $b \gets M_{i,j-1}$
  \State $c \gets M_{i-1,j-1}$
  \State $d \gets M_{i-1,j}$
  \State $e \gets M_{i-1,j+1}$
  \State $ M_{i,j} \gets L(a,b,c,d,e)$
  \EndFor \\

  \Return $M$
 
\end{algorithmic}
\end{algorithm} 

\section{The Lookup Table}
The lookup table $L$ can be explained by three invariants. No choice will ever violate one of the invariants and when for a given context neither choice would violate an invariant, a random choice is made. Note however that due to the fact that the context variables might fall outside the maze, depending on the auxiliary values that are used to replace them, effectively some invariants may be violated. 
The invariants are:

\begin{enumerate}[label=\protect\circled{\arabic*}]
    \item No $2\times 2$ squares of the same type are allowed
    \item No wall or path is allowed to start or end with thickness one
    \item Every path in any given line must be connected to a path in the next line 
\end{enumerate}

In invariant  \circled{2} ``to start or end'' is meant in a vertical sense i.e. top-down. Invariant  \circled{3} is enforced by ensuring that every 0 (path) in a line must either have a 0 in the line directly below it or a 0 to it's right. Since the lines are being built left to right at some point a continuous sequence of 0s would hit the wall on the right and then be connected to the line below. Here, it is important to remember that due to the fact that some of the context variables might fall outside of the maze it is possible that effectively the invariant is violated on either wall, depending on the auxiliary choice of values for $a$, $b$ and $e$. In the algorithm's Atari implementation the auxiliary choice for $a$ and $b$ is $0,1$ and for $c$ and $e$ it is \textsf{random}. This is why, from time to time, there are non connected paths in the game. A choice of $a = b = e = 1$ would enforce full connectivity since then the context would always match the lookup table. In the mazes that were used for the game, connectivity of all paths was not desirable thus it was not enforced. In other words: the lookup table always enforces the invariants; it might just be the case that the context in the lookup table and the real maze context differ. \\
To see how invariants \circled{1} \circled{2} \circled{3} explain the lookup table it suffices to express them in terms of the variables $a,b,c,d,e$. We can formulate all the rules stemming from the invariants as follows (The notation $010\star\star \rightarrow 1$ would mean ``no matter what $d$ and $e$ are, when $ a = 0 \land b = 1 \land c = 0$ the choice $x = 1$ is made''): 
\begin{itemize}
    \item Invariant  \circled{1} gives:\\
    \begin{tabular}{ccccccc}
       $\star$  &0  &0 &0 & $\star$& $\rightarrow$ & 1\\
       $\star$  &1  &1 &1 & $\star$& $\rightarrow$ & 0\\
    \end{tabular}
    \item Invariant  \circled{2} gives:\\
    \begin{tabular}{ccccccc}
       $\star$  &$\star$  &0 &1 & 0&$\rightarrow$ & 1\\
       $\star$  &$\star$  &1 &0 & 1&$\rightarrow$ & 0\\
       0  &1  &0 &$\star$ & $\star$&$\rightarrow$ & 1\\
       1  &0  &1 &$\star$ & $\star$&$\rightarrow$ & 0\\
    \end{tabular}
    \item Invariant  \circled{3} gives:\\
    \begin{tabular}{cccccccc}
       $\star$  &1  &0 &0 & 1& $\rightarrow$ & 0&\\
       $\star$  &$\star$  &1 &0 & 1& $\rightarrow$ & 0& (already a rule)\\
    \end{tabular}    
\end{itemize} 

Note that to enforce invariant  \circled{3}, it is sufficient to enforce  \circled{3} only for the paths that fall into variable $d$ since every $0$ in the row above will always fall into the role of variable $d$ once. 

Our claim that all entries in the lookup table can be explained by the invariants would predict that all remaining entries are $\mbox{\textsf{random}}$. This is almost the case. We still have one rule that seemingly cannot be explained, namely $00100 \rightarrow 0$. That remaining entry is explained by the remarkable fact that the invariants almost never contradict themselves. Only in one case they run into a contradiction, namely in the context $01001$ where invariant  \circled{2} demands a $1$ while invariant  \circled{3} demands a $0$. The solution is elegant: the choice $1$ is made and the rule $\star 0100 \rightarrow 0$ is added (which explains the remaining entry since $10100 \rightarrow 0$ already is a rule). This solves the problem that the path at variable $d$ would be blocked (and thus violate invariant  \circled{3}) if the block to the left of variable $c$ was a wall. The added rule enforces that such a situation will never arise (because for this to happen in the step before the context would have been $\star0100$ and a $1$ would have had been placed). It will ensure that whenever a context like $01001$ is met, putting down a wall will not disconnect the path since the $0$ at variable $d$ will always be connected to the $0$ at $a$.

Now we can also see how the predefined values for variables $a,b,e$ play a role when they fall outside the maze. In the game's implementation the choice $a = 0, b = 1$ was made. This can lead to situations where a path is only being connected to the imaginary $0$ that lies outside of the maze. Thus effectively a dead-end at the wall is created. A choice $a = b = e = 1$ will lead to no such dead-ends.

%\pagebreak

\section{The Lookup Table for 3 Dimensions}

To extend the method to 3 Dimensions we decided to maintain the same invariants. A new context had to be chosen to accommodate for the third dimension. Note that more variables will lead to more rows in the lookup table that need to be filled. We propose a 10-variable context leading to an increase from 32 to 1024 possible decisions. It is of the following form ($x$ is the new block that will be placed):
\pagebreak

\tdplotsetmaincoords{70}{120} % set viewpoint 
\tdplotsetrotatedcoords{0}{0}{0} %<- rotate around (z,y,z)

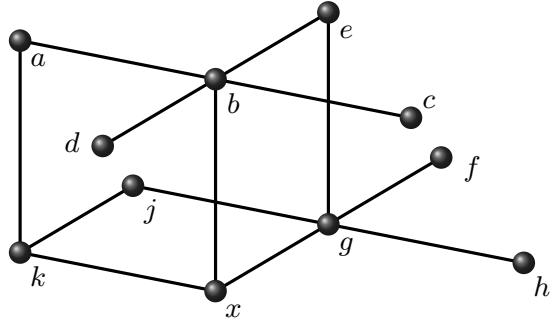
\begin{figure}
    \centering
\begin{tikzpicture}[scale=3,tdplot_rotated_coords,
                    rotated axis/.style={->,purple,ultra thick},
                    blackBall/.style={ball color = black!80},
                    borderBall/.style={ball color = white,opacity=.25}, 
                    very thick]

\draw (0,0,0) -- (0,2,0);
\draw (0,0,0) -- (1,0,0);
\draw (1,0,0) -- (1,1,0);
\draw (1,0,0) -- (1,0,1);
\draw (1,1,0) -- (0,1,0);
\draw (1,1,0) -- (1,1,1);
\draw (1,1,1) -- (1,0,1);
\draw (1,1,1) -- (0,1,1);
\draw (0,1,1) -- (0,1,0);
\draw (0,1,0) -- (-1,1,0);
\draw (1,1,1) -- (1,2,1);
\draw (1,1,1) -- (2,1,1);

\shade[rotated axis,blackBall] (0,0,0) circle (0.05cm); % j
\shade[rotated axis,blackBall] (1,0,1) circle (0.05cm); % a
\shade[rotated axis,blackBall] (1,2,1) circle (0.05cm); % c
\shade[rotated axis,blackBall] (0,1,1) circle (0.05cm); % e
\shade[rotated axis,blackBall] (2,1,1) circle (0.05cm); % d
\shade[rotated axis,blackBall] (1,1,0) circle (0.05cm); % x
\shade[rotated axis,blackBall] (1,0,0) circle (0.05cm); % k
\shade[rotated axis,blackBall] (0,1,0) circle (0.05cm); % g
\shade[rotated axis,blackBall] (-1,1,0) circle (0.05cm); % f
\shade[rotated axis,blackBall] (0,2,0) circle (0.05cm); % h
\shade[rotated axis,blackBall] (1,1,1) circle (0.05cm); % b
\draw (0.1,0.15,0.05) node[scale=1.2, below] {$j$}; 
\draw (0.1,1.15,1.05) node[scale=1.2, below] {$e$}; 
\draw (0.1,1.15,0.05) node[scale=1.2, below] {$g$};
\draw (-1.1,1.1,0.05) node[scale=1.2, below] {$f$};
\draw (0.1,2.15,0.05) node[scale=1.2, below] {$h$};
\draw (1.1,1.15,1.05) node[scale=1.2, below] {$b$};   
\draw (1.1,1.15,0.05) node[scale=1.2, below] {$x$}; 
\draw (2.1,1,1.05) node[scale=1.2, left] {$d$};
\draw (1.1,0.15,1.05) node[scale=1.2, below] {$a$};
\draw (1.1,0.15,0.05) node[scale=1.2, below] {$k$};
\draw (1.1,2.15,1.05) node[scale=1.2, above] {$c$};
\end{tikzpicture}
    \caption{3 Dimensional Maze Generation Context}
    \label{fig:my_label}
\end{figure}

The following rules are applied:

\begin{itemize}
    \item Invariant \circled{1} gives:\\
    \begin{tabular}{cccccccccccc}
       0 &0 &$\star$ &$\star$ &$\star$ &$\star$ &$\star$ &$\star$ &$\star$ &0 & $\rightarrow$ & 1\\
       1 &1 &$\star$ &$\star$ &$\star$ &$\star$ &$\star$ &$\star$ &$\star$ &1 & $\rightarrow$ & 0\\
       $\star$ &$\star$ &$\star$ &$\star$ &$\star$ &$\star$ &0 &$\star$ &0 &0 & $\rightarrow$ & 1\\
       $\star$ &$\star$ &$\star$ &$\star$ &$\star$ &$\star$ &1 &$\star$ &1 &1 & $\rightarrow$ & 0\\
       $\star$ &0 &$\star$ &$\star$ &0 &$\star$ &0 &$\star$ &$\star$ &$\star$ & $\rightarrow$ & 1\\
       $\star$ &1 &$\star$ &$\star$ &1 &$\star$ &1 &$\star$ &$\star$ &$\star$ & $\rightarrow$ & 0\\
    \end{tabular}
    \item Invariant \circled{2} gives:\\
    \begin{tabular}{cccccccccccc}
       0 &1 &0 &0 &0 &$\star$ &$\star$ &$\star$ &$\star$ &$\star$ & $\rightarrow$ & 1\\
       1 &0 &1 &1 &1 &$\star$ &$\star$ &$\star$ &$\star$ &$\star$ & $\rightarrow$ & 0\\
       $\star$ &$\star$ &$\star$ &$\star$ &0 &0 &1 &0 &0 &$\star$ & $\rightarrow$ & 1\\
       $\star$ &$\star$ &$\star$ &$\star$ &1 &1 &0 &1 &1 &$\star$ & $\rightarrow$ & 0\\
    \end{tabular}
    \item Invariant \circled{3} gives:\\
    \begin{tabular}{ccccccccccccc}
       0 &0 &1 &1 &1 &$\star$ &$\star$ &$\star$ &$\star$ &1 & $\rightarrow$ & 0\\
       1 &0 &1 &1 &0 &$\star$ &1 &$\star$ &$\star$ &$\star$ & $\rightarrow$ & 0\\
       0 &0 &1 &1 &0 &$\star$ &1 &$\star$ &$\star$ &1 & $\rightarrow$ & 0\\
    \end{tabular}    
\end{itemize}

Just as with the bi-dimensional invariants, conflicting situations can occur. I.e. in case of a context of the form $11\star\star001001$ invariant \circled{1} demands a $0$ since otherwise a $2 \times 2$ square of walls is formed along the variables $a, b, k$ and the new block. But at the same time invariant \circled{2} demands a $1$ since otherwise the $1$ at variable $g$ will form the start of a wall with thickness one. Just as it was done for the 2D lookup table, new rules are added to prevent these contexts from appearing. Thus the following new rules are added:

\begin{itemize}
    \item To prevent conflicts \begin{tabular}{ccccccccccc}
       1 &1 &$\star$ &$\star$ &0 &0 &1 &0 &0 &1
    \end{tabular}: \\
    \begin{tabular}{cccccccccccc} $\star$ &1 &1 &$\star$ &$\star$ &$\star$ &0 &1 &$\star$ &$\star$ & $\rightarrow$ & 0 \end{tabular}
    
    \item To prevent conflicts \begin{tabular}{ccccccccccc}
       0 &0 &$\star$ &$\star$ &1 &1 &0 &1 &1 &0
    \end{tabular}: \\
    \begin{tabular}{cccccccccccc} $\star$ &0 &0 &$\star$ &$\star$ &$\star$ &1 &0 &$\star$ &$\star$ & $\rightarrow$ & 1 \end{tabular}
    
    \item To prevent conflicts \begin{tabular}{ccccccccccc}
       0 &1 &0 &0 &0 &$\star$ &1 &$\star$ &1 &1
    \end{tabular}: \\
    \begin{tabular}{cccccccccccc} $\star$ &0 &1 &$\star$ &$\star$ &$\star$ &1 &1 &$\star$ &$\star$ & $\rightarrow$ & 0 \end{tabular}

    \item To prevent conflicts \begin{tabular}{ccccccccccc}
       1 &0 &1 &1 &1 &$\star$ &0 &$\star$ &0 &0
    \end{tabular}: \\
    \begin{tabular}{cccccccccccc} $\star$ &1 &0 &$\star$ &$\star$ &$\star$ &0 &0 &$\star$ &$\star$ & $\rightarrow$ & 1 \end{tabular}
    
    \item To prevent conflicts \begin{tabular}{ccccccccccc}
       $\star$ &0 &1 &1 &0 &0 &1 &0 &0 &$\star$
    \end{tabular}: \\
    \begin{tabular}{cccccccccccc} $\star$ &$\star$ &0 &1 &1 &$\star$ &0 &1 &$\star$ &$\star$ & $\rightarrow$ & 0 \end{tabular}
\end{itemize}

Naturally more dimensions lead to a bigger context which, in turn, calls for more rules. Therefore we also see more conflicts between the rules. In the bi-dimensional case there was one conflict in 32 possible contexts (roughly $3\%$) while in the three dimensional case we have 28 conflicts in 1024 possible contexts (also roughly $3\%$). Although in the higher dimensional case we have the added complexity that the conflict preventing rules themselves run into conflicts. Thus only $75\%$ of the conflicts could be prevented which leads to roughly $0.7\%$ of contexts in which a choice will violate an invariant. This number could be further minimized by increasing the size of the context or the maze generation could be re-run with a different random seed.   

\begin{figure}
    \centering
    \includegraphics[width=0.7\textwidth]{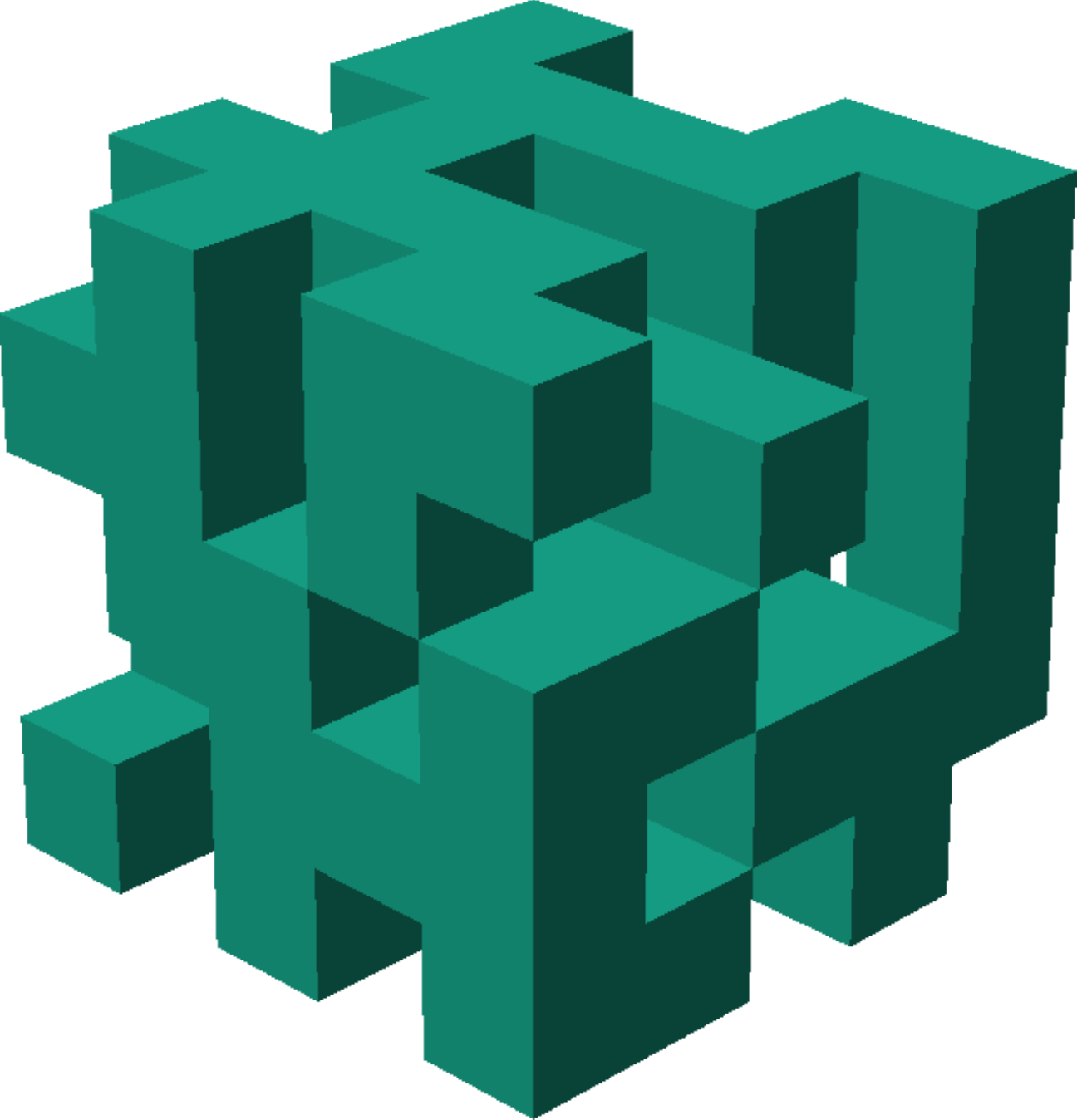}
    \caption{$ 5 \times 5 \times 5$ example maze}
    \label{fig:cube3}
\end{figure}

\begin{figure}
    \centering
    \includegraphics[width=0.7\textwidth]{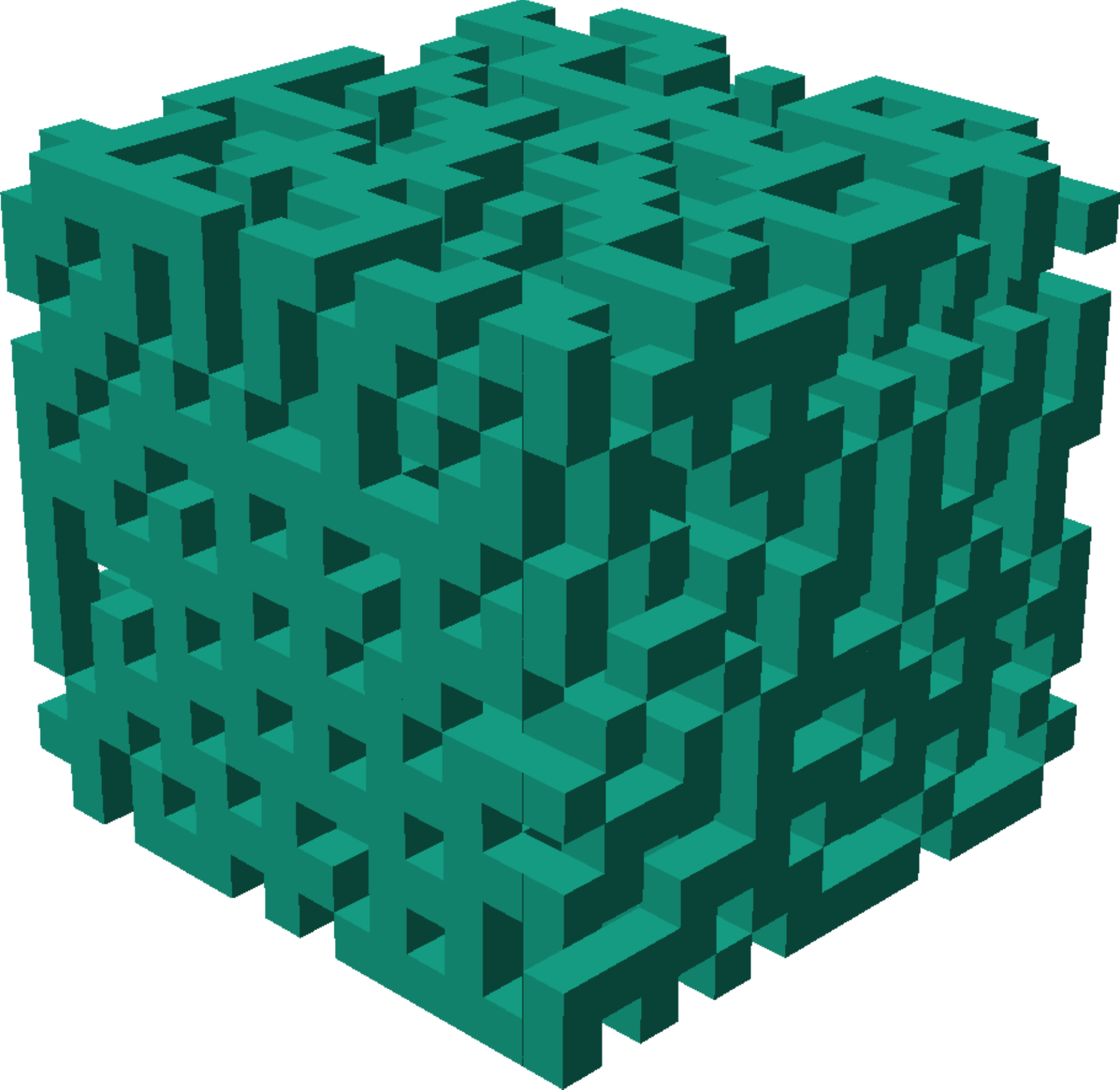}
    \caption{$ 15 \times 15 \times 15$ example maze}
    \label{fig:cube4}
\end{figure}

\begin{figure}
    \centering
    \includegraphics[width=0.7\textwidth]{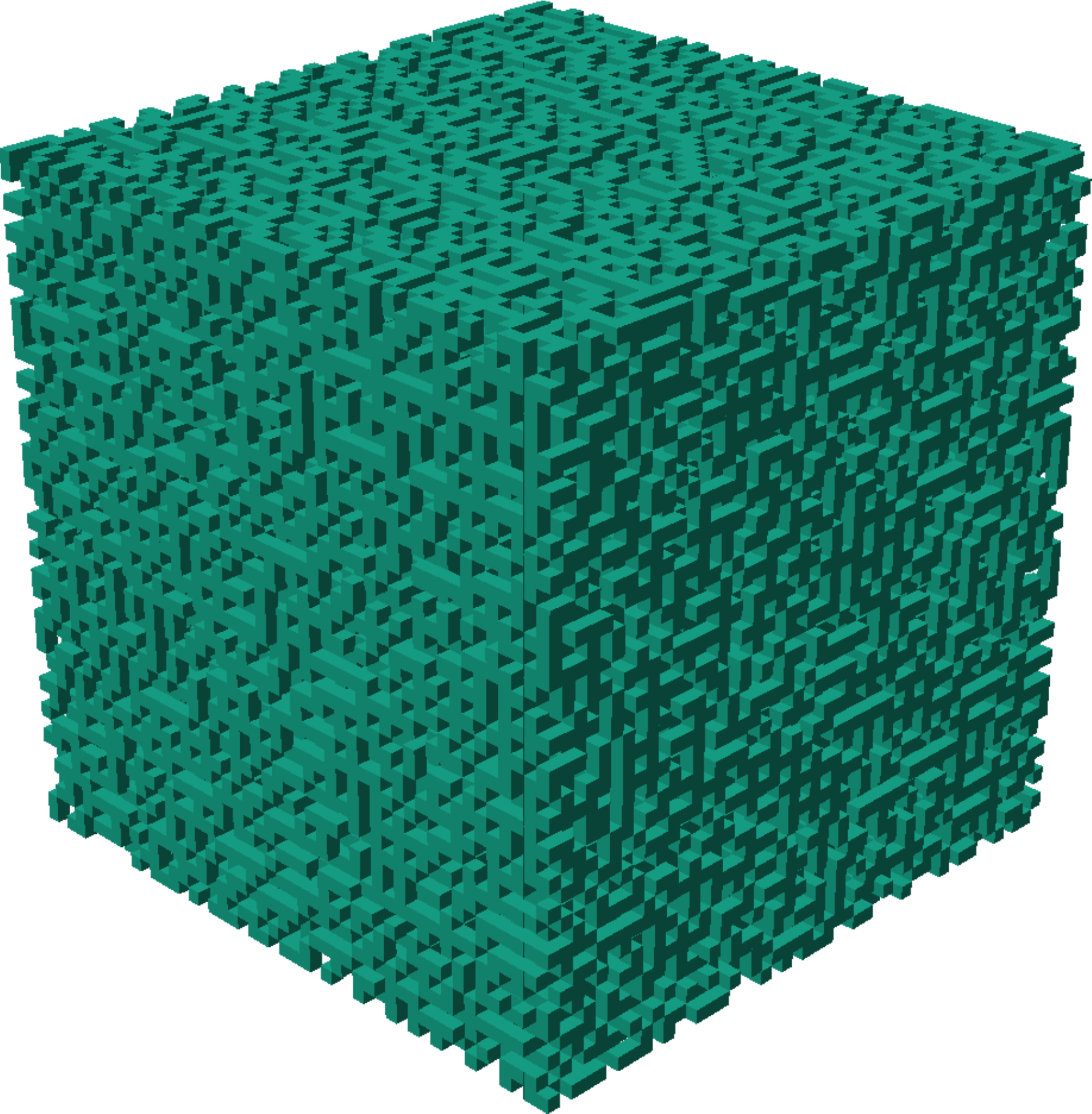}
    \caption{$ 50 \times 50 \times 50$ example maze}
    \label{fig:cube5}
\end{figure}

\bibliographystyle{acm}
\bibliography{mers.bib}

\clearpage
\appendix

%\end{comment}
\end{document}